\documentclass[conference]{IEEEtran}
\IEEEoverridecommandlockouts

\usepackage{cite}
\usepackage{amsmath,amssymb,amsfonts}
\usepackage{algorithmic}
\usepackage{graphicx}
\usepackage{textcomp}
\usepackage{xcolor}
\usepackage{booktabs}
\usepackage{array}
\usepackage[hidelinks]{hyperref}

\begin{document}

\title{\text{D}iagnosing \text{O}verhead in \text{D}ispatch \text{O}perations: \text{C}ross-architecture \text{O}bservatory}

\author{%
\IEEEauthorblockN{Bole Ma, Jan Eitzinger, Harald K\"ostler, and Gerhard Wellein}
\IEEEauthorblockA{\textit{Erlangen National High-Performance Computing Center (NHR@FAU)} \\
Erlangen, Germany \\
\{bole.ma, jan.eitzinger, harald.koestler, gerhard.wellein\}@fau.de}
}


\maketitle

\begin{abstract}
AlltoAll dispatch is the dominant bottleneck of MoE expert parallelism, and the interconnect community has responded with four families of mitigations: predictive sample placement, adaptive expert relayout, hierarchical collectives, and EP-aware topology. All four rest on two assumptions about the workload: that routing imbalance is correctable by the system layer, and that the mock-token benchmarks evaluating them faithfully represent production routing. We introduce DODOCO to test both, instrumenting five open MoE checkpoints that span today's sequence-mixer designs (MHA, MLA, GQA, Gated DeltaNet and Mamba-2 SSM) under a factorial grid of six data conditions and a matched expert-parallelism scan on H100 clusters.

Both assumptions fail. Scaling EP leaves per-expert load concentration essentially unchanged: the straggler is intrinsic to the routing decision the model makes, not to how its experts land on ranks. Mock tokens overestimate routing imbalance by up to a factor of $2.35$, and the error is a level offset rather than a trend: it stays flat across a $32\times$ batch-size sweep. Skewing the synthetic distribution toward realism (Zipf) widens the gap instead of closing it.

A third pattern organizes the results: the architectures separate into a data-resilient band (MHA, Mamba-2), whose routing approaches uniform on real text, and a persistently concentrated band (MLA, GDN), with GQA intermediate. These bands, not the EP degree or the mock-data profile, are the right workload input to AlltoAll-aware interconnect and dispatch design.
\end{abstract}

\begin{IEEEkeywords}
Mixture-of-Experts, AlltoAll, expert parallelism, communication characterization, load imbalance, interconnect
\end{IEEEkeywords}

\section{Introduction}
\label{sec:intro}

A frontier MoE training step spends roughly half of its forward-pass time inside two AlltoAll calls~\cite{bernadskiy2025lightmatter}. Four families of system-level responses have emerged: \emph{predictive} placement that exploits routing locality across iterations~\cite{netmoe2025}; \emph{adaptive} expert relayout that responds to measured load~\cite{laermoe2026}; \emph{hierarchical} collectives that aggregate AlltoAll traffic in stages~\cite{tran2024ohio,aderholdt2024ucc}; and \emph{topological} responses that redesign the network around EP traffic~\cite{wang2024railonly,bernadskiy2025lightmatter}. These are not in conflict with each other (they address different layers of the same stack), but they share two assumptions that, to our knowledge, have not been examined head-on.

\textbf{Assumption A (correctability).} Routing imbalance is a workload property a sufficiently clever system layer can reduce.

\textbf{Assumption B (benchmark validity).} How systems are evaluated (almost always with mock or randomly-generated token sequences, because real data is awkward to provision in a comm benchmark; HeterMoE~\cite{pan2025hetermoe} evaluates explicitly on random tensors, Tutel~\cite{hwang2023tutel} uses skewed-Zipf synthetic dispatch patterns) is a faithful proxy for production routing.

Our work, DODOCO, tests both, empirically with the following three concrete questions organizing the paper.

\begin{itemize}
\item \emph{Q1.} If a scheduler doubles EP from 16 to 32, does the per-rank load distribution become more uniform? Can EP scaling alone reduce the straggler?
\item \emph{Q2.} If a benchmark uses mock tokens (uniform-random IDs), how much does the measured routing imbalance differ from what the same model produces on real text?
\item \emph{Q3.} Does the answer to Q1 or Q2 depend on the attention mechanism, or are the five attention designs deployed across DeepSeek, Qwen, and Nemotron families interchangeable from a communication-characterization standpoint?
\end{itemize}

Our core experiment is a $5{\times}6$ factorial study (every combination of five architectures and six data conditions, 30 cells total) on four H100 nodes (16 EP ranks, NVLink intra-node, InfiniBand inter-node), with a separate controlled EP scan (up to eight nodes) and a separate global-batch-size sweep. We use five MoE checkpoints released by their authors with their as-deployed routing configurations: DeepSeek-V2-Lite~\cite{deepseekv2} (Multi-head Latent Attention, MLA), DeepSeek-MoE-16B~\cite{dai2024deepseekmoe} (standard Multi-Head Attention, MHA), Qwen3-30B-A3B~\cite{yang2025qwen3} (Grouped-Query Attention, GQA), Nemotron-30B-A3B~\cite{nvidia2025nemotron} (Mamba-2 state-space sequence mixer), and Qwen3.5-35B-A3B (Gated Delta Networks, GDN; a linear-attention variant). The MLA and MHA models share expert count (64), top-$k$ (6), gating, and load-balancing strategy and differ only in attention mechanism, a controlled pair that lets us isolate attention from MoE configuration.

The findings can be stated briefly.

\emph{Q1: No.} Within the EP range each architecture could be scanned, the per-expert max/mean token ratio is essentially flat. The MLA/MHA controlled pair varies by $5.0\%$ and $2.8\%$ across the full EP=$\{4,8,16,32\}$ range; the other three architectures vary by $0.3\%$ (GDN), $4.4\%$ (Mamba-2), and $4.4\%$ (GQA) across EP=$\{8,16,32\}$. A $\leq 5\%$ change is well below what any system-side intervention is designed to recover, and it does not vary in a monotonic direction across the five architectures: the EP degree is not the lever for routing balance. Scaling EP changes which rank holds which expert, not which experts the model sends tokens to, so the per-rank skew the AlltoAll actually pays is bounded below by this EP-invariant concentration. We are not aware of a prior matched-window measurement of MoE routing imbalance as a function of EP degree across multiple sequence-mixer architectures; existing characterizations profile at a fixed parallelization~\cite{nie2023flexmoe}, exclude EP~\cite{anthony2024demystifying}, or report a single EP point per model~\cite{laser,latentproto2025}.

\emph{Q2: Substantially.} Mock-token inputs overestimate routing Gini by up to $2.35\times$, and the overestimate is a constant-factor level error rather than a trend: on the architecture-controlled MLA/MHA pair, a three-seed re-measurement shows per-rank imbalance flat to within $3\%$ across a $32\times$ batch-size sweep under mock and real text alike (\S\ref{sec:mock}). A skewed-Zipf synthetic variant, the natural next guess for a cheap realistic proxy, overestimates real-text imbalance by $1.2\times$ to $3.9\times$, exceeding even uniform mock for every architecture. The largest mock-to-real gap, on Nemotron, reaches a factor of $47\times$ when expressed as fitted Dirichlet $\alpha$ (a finer-grained uniformity metric than Gini), meaning a profile calibrated on mock tokens places the operating point in a qualitatively different region of the uniformity simplex. Recent benchmark work~\cite{moeinferencebench2025} flagged the question of whether findings from synthetic data transfer to real language distributions as open; our 5$\times$6 matrix is a direct answer for the routing-imbalance dimension.

\emph{Q3: Yes.} Across data conditions, the five architectures separate by how much real data helps. MHA and Mamba-2 reach the lowest per-rank Gini under real text ($0.105$ and $0.150$); MLA and GDN stay above $0.24$ even on their best English-text condition and rise to $0.29$--$0.38$ on mock. GQA is intermediate: its mock-to-real improvement is real but smaller ($1.4\times$ vs $\geq 2\times$ for MHA/Mamba-2), and its wikitext Gini ($0.240$) sits next to MLA's ($0.245$) in absolute terms. We call the cleanest two ends \emph{data-resilient} and \emph{persistently concentrated} and treat GQA as a mixed case that the paper labels explicitly each time it appears. The architectural coverage here is broader than prior cross-model MoE routing characterizations~\cite{laser,latentproto2025}, none of which cover MLA, Mamba-2 SSM, or linear-attention GDN. A per-layer view sharpens the picture further: the data-resilient pair starts each MoE block near-uniformly and concentrates with depth, while the persistent pair is concentrated from the very first MoE layer.

\textbf{Our contribution.} Anthony et al.~\cite{anthony2024demystifying} characterized DP, TP, and PP for dense transformers at HotI 2024 and explicitly left expert parallelism as future work; a HotI 2025 follow-up~\cite{xu2025characterizing} extends that series to distributed LLM inference. We provide the MoE expert-parallel characterization. Our point is not that interconnect design is misguided; it is that the workload model practitioners reach for when designing or evaluating it is at best a constant-factor approximation. The two-class taxonomy, not the EP degree or the mock-data profile, is the more useful input to system design.

\textbf{What we do not claim.} We do not measure end-to-end training speedup from any system-side intervention; we characterize routing decisions and the AlltoAll-relevant statistics they produce. We do not claim that imbalance is unimportant: it is large, and Section~\ref{sec:discussion} discusses what is and is not addressable by the named systems above. And we do not claim that mock-data evaluation has been worthless; we claim it has been miscalibrated.

\section{Background}
\label{sec:bg}

Readers familiar with MoE expert parallelism can skim \S\ref{sec:bg:moe-ep} and focus on \S\ref{sec:bg:archs}; readers familiar with recent attention/sequence-mixer designs should do the reverse.

\subsection{MoE Expert Parallelism and Its AlltoAll}
\label{sec:bg:moe-ep}

In an MoE block, a lightweight router scores each input token against $E$ expert FFNs and dispatches the token to its top-$k$ experts. Under expert parallelism with $P$ ranks, the $E$ experts are partitioned across ranks; each rank then sends its locally-produced tokens to the rank that holds their assigned expert and receives in turn the tokens routed to the experts it holds. Token dispatch and combine therefore become two AlltoAll collectives per MoE block, parameterized by a \emph{send\_counts} matrix $S \in \mathbb{N}^{P{\times}P}$ where $S_{ij}$ is the number of tokens rank $i$ sends to rank $j$.

The bottleneck behavior of AlltoAll-with-skew is standard. The collective completes only when every rank has received all of its incoming tokens, so completion time tracks the \emph{maximum column sum}, not the mean. When the routing distribution is uniform, $\max_j \sum_i S_{ij} \approx \mathbb{E}[\sum_i S_{ij}]$ and the bandwidth-bound model gives the expected lower bound on time. When routing concentrates, $\max_j \sum_i S_{ij}$ can be several times the mean, and every other rank waits. This is the same straggler logic that motivates load-balancing losses~\cite{fedus2022switch}, capacity factors~\cite{lepikhin2021gshard}, dynamic shadowing~\cite{he2022fastermoe}, adaptive collectives~\cite{hwang2023tutel,tran2024ohio}, and predictive placement~\cite{netmoe2025,laermoe2026}.

We use three scalar summaries of $S$, none of them new but each with a different sensitivity. Let $c_j = \sum_i S_{ij}$ be the per-rank receive count and $e_\ell$ the per-expert receive count.

The \emph{per-expert max/mean ratio} $\max_\ell e_\ell / \overline{e}$ is the primary statistic for Q1. It captures the routing decision in a form that EP cannot influence: EP scaling only changes the expert-to-rank bijection, not which experts the router picks. The per-rank max/mean that bandwidth-bound AlltoAll completion tracks directly is bounded below by the per-expert ratio scaled by $P/E$: if a single expert receives several times the mean expert load, no remapping of experts across ranks can reduce that rank's share. A per-expert ratio that is invariant across the EP scan therefore tells us EP scaling cannot reduce AlltoAll skew, regardless of how the bijection is chosen.

The \emph{Gini coefficient}~\cite{laser,latentproto2025,yi2026threephase} of per-rank receive counts, with sorted $c_{(1)} \le \dots \le c_{(P)}$,
\begin{equation}
G \;=\; \frac{\sum_{j=1}^{P}\bigl(2j - P - 1\bigr)\,c_{(j)}}{P \sum_j c_j},
\label{eq:gini}
\end{equation}
ranges from 0 (perfect balance) to 1 (one rank absorbs all). We use per-rank Gini for the cross-condition comparisons in \S\ref{sec:mock} and \S\ref{sec:tax}, where both the routing decision and the bijection contribute and a single per-rank summary is the relevant operational quantity.

The fitted symmetric \emph{Dirichlet concentration} $\alpha$ is a complementary metric we use specifically for the near-uniform regime that Gini compresses. We treat the per-rank receive proportions $p_j = c_j / \sum_{j'} c_{j'}$ as a draw from a symmetric Dirichlet and report the moment-matched $\alpha$. The intuition is that $\alpha$ measures how tightly the per-rank distribution clusters around the perfectly-uniform point: $\alpha \to \infty$ means concentrated at uniform, $\alpha \to 0$ means concentrated at a single-rank vertex. Where two cells that both look ``mostly balanced'' compress into nearby Gini values of, say, $0.05$ and $0.10$, the same cells can sit at $\alpha = 40$ vs $\alpha = 8$, a $5\times$ ratio that resolves the contrast that Gini hides. We report $\alpha$ alongside Gini wherever the latter is small enough to compress meaningful differences. Section~\ref{sec:ep} validates that per-rank Gini predicts measured dispatch P50 within each architecture with $r \geq 0.99$ and P99 with $r \geq 0.76$, so the abstract metric tracks operational AlltoAll latency in our setting.

\subsection{The Attention Mechanisms We Cover}
\label{sec:bg:archs}

We span five sequence-mixer designs that together account for nearly all currently-deployed MoE checkpoints, with two softmax-attention variants (MHA, MLA) forming an attention-controlled pair and one model each for the other three:

\begin{itemize}
\item \textbf{Multi-Head Attention (MHA)}: DeepSeek-MoE-16B~\cite{dai2024deepseekmoe}. The classical transformer attention.

\item \textbf{Multi-head Latent Attention (MLA)}: DeepSeek-V2-Lite~\cite{deepseekv2}. A rank-512 compression of the key-value cache, designed primarily to reduce KV memory; the compression also constrains the hidden-state geometry the router sees. With MHA above, shares everything but the attention mechanism, isolating the attention effect.

\item \textbf{Grouped-Query Attention (GQA)}: Qwen3-30B-A3B~\cite{yang2025qwen3}. Heads share keys and values in groups, reducing KV memory without compressing the head dimension.

\item \textbf{State Space Model (Mamba-2)}: Nemotron-30B-A3B~\cite{nvidia2025nemotron}. A \emph{hybrid} architecture: 23 Mamba-2 layers and 6 GQA attention layers interleaved with 23 MoE blocks. Uses sigmoid routing with an additive expert-bias term on top of an auxiliary loss; the other four models use softmax routing with the auxiliary loss alone.

\item \textbf{Gated Delta Networks (GDN)}: Qwen3.5-35B-A3B. Also a \emph{hybrid}: a 3:1 repeating pattern of three GDN (linear-attention-with-gated-delta) layers and one full-softmax attention layer, repeated 10 times for 40 total layers, each with MoE.
\end{itemize}

Expert counts and top-$k$ for all five are in Table~\ref{tab:models}.

What these architectures have in common, from a communication standpoint, is that they all reduce a sequence of token states to a single fixed-dimensional input per token at the router. What they do differently is how compressed, how diverse, and how sequentially-coherent that router input ends up being. Section~\ref{sec:tax} returns to this point with measurements.

\section{Methodology}
\label{sec:method}

\subsection{Hardware, Software, and Scope}

All measurements run on the Helma supercomputer at NHR@FAU: four H100 GPUs per node (94\,GB HBM2e each), NVLink intra-node and NDR200 InfiniBand inter-node (four ConnectX-7 HCAs per node bound to NCCL in rail-optimal one-HCA-per-GPU mapping, $\texttt{NCCL\_IB\_GID\_INDEX}{=}3$). The software stack is PyTorch with NVIDIA NeMo 26.02 and Megatron-Core 0.16.1, executed inside an Apptainer container; sequence length is 4096 and precision is bf16 (the released checkpoints' native dtype). Unless otherwise noted, EP=16 (one expert-parallel rank per GPU; one tensor-parallel rank). The factorial $5{\times}6$ block uses GBS=16 with 5 warmup and 10 measurement iterations on pretrained weights; the EP scan and the GBS sweep use larger windows described below. All data conditions are reproducible from public sources (HuggingFace datasets cited inline below) using standard Megatron-Core tokenization; the only project-specific code is a small instrumentation patch on \texttt{MoEAlltoAllTokenDispatcher} that records \texttt{input\_splits} per layer.

We use each model's released checkpoint with its as-deployed routing configuration (Table~\ref{tab:models}; Appendix~\ref{app:notation} unpacks the load-balancing notation for readers outside the LLM training literature). All five models use aux-loss-based balancing; aux-loss-free routing (DeepSeek-V3 bias-shift, Ling-style) is explicitly outside this matrix's scope. The as-deployed choice is the ecologically relevant comparison (it captures models as they would actually be deployed and benchmarked), and it means that Nemotron's routing differs from the other four along two axes simultaneously (sigmoid scoring, additive expert-bias term); \S\ref{sec:tax} returns to this confound with a targeted ablation.

\begin{table}[t]
\caption{Routing configuration of the five MoE checkpoints. $E$: expert count, $k$: top-$k$, coef.: aux-loss coefficient. Notation defined in the text.}
\label{tab:models}
\small
\centerline{\begin{tabular}{lcccll}
\toprule
Model & Attn & $E$ & $k$ & Score / LB strategy & coef. \\
\midrule
DS-MoE   & MHA     & 64  & 6 & softmax + seq-aux       & $10^{-3}$ \\
DSv2-Lite& MLA     & 64  & 6 & softmax + seq-aux       & $10^{-3}$ \\
Qwen3    & GQA     & 128 & 8 & softmax + aux           & $10^{-3}$ \\
Nemotron & Mamba-2 & 128 & 8 & sigmoid + seq-aux + bias & $10^{-4}$ \\
Qwen3.5  & GDN     & 128 & 8 & softmax + aux           & $10^{-3}$ \\
\bottomrule
\end{tabular}}
\end{table}

\subsection{The Six Data Conditions}
\label{sec:method:data}

The six conditions form a gradient that isolates independent properties of a token stream:

\begin{enumerate}
\item \textbf{Mock} -- uniform-random token IDs. No frequency structure, no sequential structure, no learned embedding meaning. The de facto default for MoE communication benchmarks.

\item \textbf{Shuffled} -- real wikitext tokens, then a global permutation across all documents. Preserves the token-frequency distribution of real text; destroys all sequential structure.

\item \textbf{Remapped} -- real wikitext sequences with a fixed random permutation applied to vocabulary IDs. Preserves sequential structure (the same $n$-gram patterns appear in index space); destroys learned token--embedding associations.

\item \textbf{Romansh} -- real text in Rumantsch Grischun, a Romance language with $\sim$60{,}000 speakers, sourced from \texttt{swiss-ai/apertus-pretrain-romansh}~\cite{apertus2025}. Real linguistic structure in a language absent from standard pretraining corpora.

\item \textbf{Opus} -- post-cutoff English reasoning traces. Familiar language structure, content the models cannot have seen.

\item \textbf{Wikitext} -- English Wikipedia (wikitext-103~\cite{merity2017wikitext}). Familiar language structure, content very likely included in pretraining.
\end{enumerate}

The shuffled--mock comparison isolates token frequency. The remapped--wikitext comparison isolates learned embeddings. The Romansh--Opus comparison isolates language familiarity. The Opus--wikitext comparison isolates seen-vs-unseen content. Conditions 4--6 are real text streamed from pre-tokenized Megatron binaries.

\subsection{Instrumentation}

We log per-dispatch \texttt{input\_splits} from \texttt{MoEAlltoAllTokenDispatcher} on every EP rank and at every MoE layer, recovering the full $S$ matrix for each step. Timing is collected on every rank with CUDA events; we report system P99 as $\max$ over ranks (the relevant statistic, since the slowest rank determines AlltoAll completion).

\subsection{EP and Batch-Size Scans}
\label{sec:method:scans}

The factorial block reports Gini, $\alpha$, and max/mean at EP=16, GBS=16. For Q1 (EP scaling), we run a separate \emph{matched-window} scan: the same global batch size (GBS=32) and the same iteration budget (50 warmup + 200 measurement) at every EP point in $\{4, 8, 16, 32\}$, so that any difference in measured imbalance reflects EP itself rather than data exposure or sample size. We use this configuration for all claims about how EP affects imbalance.

We complete all four EP points (4, 8, 16, 32) for the MLA/MHA pair with TP=1 (64 experts at $d{=}2048$ fit at EP=4). For GQA, Mamba-2, and GDN, EP=4 does not fit in the H100's 94\,GB below TP=4, and a fixed-EP control shows the TP=4 layout itself shifts the per-expert statistic ($+18\%$ on GQA's max/mean); we therefore scan EP=$\{8, 16, 32\}$ for those three, with TP=2 at EP=8 for Mamba-2 and GDN --- points the scan itself validates (within $2.7\%$ and $0.3\%$ of their TP=1 neighbors). The flatness claim is correspondingly scoped to EP=$\{8,16,32\}$ for these three architectures. Appendix~\ref{app:tp} gives the full TP-sensitivity controls and why closing the EP=4 gap requires larger-memory GPUs rather than more tensor parallelism.

The four EP points also span the cluster's full bandwidth hierarchy, from single-node NVLink-only AlltoAll at EP=4 to eight-node IB-dominated AlltoAll at EP=32, an $\sim 18\times$ per-GPU bandwidth transition; Appendix~\ref{app:topo} details the topology footprint of each scan point. The factorial $5{\times}6$ cells all use EP=16 (the four-node, IB-dominated configuration).

For Q2 (mock vs real), we sweep GBS $\in \{64, 256, 1024, 2048\}$ at EP=16 across all five architectures under both mock and wikitext; for the MLA/MHA pair, the mock arm of this sweep is additionally measured under three RNG seeds with per-dispatch \texttt{input\_splits} logging at every point, so batch-size effects can be separated from run-to-run variance (\S\ref{sec:gbs}).

\section{Q1: EP Scaling Does Not Reduce the Straggler}
\label{sec:ep}

\begin{figure}[t]
\centerline{\includegraphics[width=\columnwidth]{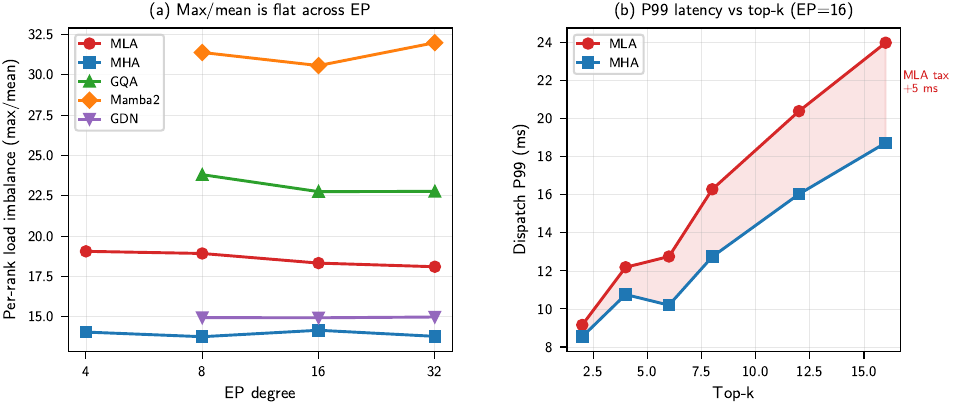}}
\caption{(a) Per-expert max/mean token ratio vs EP degree, matched windows. The metric bounds AlltoAll skew from below independently of expert-to-rank placement. Within each architecture the line is flat ($\leq 5\%$) even though the EP axis spans the cluster's NVLink$\to$IB bandwidth hierarchy (one node at EP=4 to eight IB-dominated nodes at EP=32; Appendix~\ref{app:topo}). EP=4 is omitted for the 128-expert models (TP=4 confound, Appendix~\ref{app:tp}). (b) Sensitivity contrast: dispatch P99 vs top-$k$ on the MLA/MHA pair at EP=16 (shaded band: the MLA--MHA gap). The same per-expert statistic that stays flat across EP responds cleanly when the routing workload itself (top-$k$) changes.}
\label{fig:ep_scaling}
\end{figure}

Fig.~\ref{fig:ep_scaling}(a) shows five flat lines. The architectures sit at five different operating points (Mamba-2 highest at max/mean $\sim 31$, GQA $\sim 23$, MLA $\sim 18.5$, GDN and MHA lowest at $\sim 14$--$15$), but \emph{within} each architecture the EP scan produces $\leq 5\%$ variation in max/mean: MLA $5.0\%$, MHA $2.8\%$, GQA $4.4\%$, Mamba-2 $4.4\%$, GDN $0.3\%$. The flatness matters because the EP axis spans the cluster's full bandwidth hierarchy: EP=4 is one node with NVLink only ($\sim 450$\,GB/s aggregate per GPU, no IB), EP=8 is two nodes (roughly equal NVLink and IB partners), and EP=32 is eight nodes where the AlltoAll is IB-dominated (NDR200 IB at $\sim 25$\,GB/s per GPU, an $\sim 18\times$ per-GPU bandwidth drop from EP=4). Max/mean (a per-expert statistic of the routing distribution) does not respond to that bandwidth transition because it is a property of the routing decision, which the network does not influence.

\textbf{This is a claim about \emph{skew}, not about absolute completion time.} Absolute dispatch P99 in milliseconds does scale substantially across the EP transition on Helma: on the MLA/MHA pair, P99 grows from $\sim 1.2$\,ms at EP=4 (NVLink only) to $\sim 18$\,ms at EP=32 (IB-dominated), a $\sim 15\times$ increase that tracks the per-link bandwidth drop. The skew \emph{multiplier} our work characterizes (how much slower the slowest rank is than a balanced collective would be) is EP-invariant; the \emph{baseline} the multiplier acts on (network bandwidth) is not. Bandwidth-side optimization (faster IB, photonic interconnect) reduces the baseline; skew-side optimization (placement, re-layout) would reduce the multiplier. The two are independent, and our result constrains the latter.

A metric-validation check closes the loop within a fixed EP regime: across the 30 factorial cells (all at EP=16, the four-node IB-dominated configuration), per-cell Gini predicts dispatch P99 with $r \geq 0.76$ within each architecture ($r \geq 0.99$ for P50; not shown). We do not extend the Gini-vs-P99 validation across the EP scan because per-rank Gini changes with EP for a separate aggregation reason (fewer experts per rank at high EP increases per-rank variance) and the network-bandwidth term also moves with EP; the two effects co-vary and a within-arch correlation across EP would be a confound, not a validation. Sensitivity of max/mean to genuine workload changes is instead confirmed by a separate top-$k$ sweep (Fig.~\ref{fig:ep_scaling}b): on the MLA/MHA pair at EP=16, dispatch P99 climbs from $\sim 9$\,ms at $k{=}2$ to $\sim 24$\,ms at $k{=}16$, with the MLA tax growing from $\sim 1$\,ms to $\sim 5$\,ms: the metric moves cleanly when something other than EP changes.

The interpretation is mechanical. When EP changes, the router still produces the same per-token expert assignments; it does not know how experts are distributed across ranks. What changes is the bijection between expert index and rank, and that changes which rank receives a hot expert's traffic, not how concentrated that traffic is on its destination. Once a model has decided that some experts will receive several times the average token count, no remapping of experts to ranks reduces $\max_j \sum_i S_{ij}$ by more than a small amount; doubling the number of destinations does not redistribute tokens that were assigned to a specific expert.

\textbf{What this rules out, and what it does not.} It rules out the default assumption that scaling the EP group by itself makes the per-rank load more uniform. We are not aware of a paper that claims EP degree \emph{alone} reduces routing skew; the assumption is instead implicit in systems that co-locate or migrate experts across ranks: SmartMoE co-locates hot and cold experts~\cite{zhai2023smartmoe}, NetMoE reorders samples~\cite{netmoe2025}, and LAER-MoE migrates experts during training~\cite{laermoe2026}. Our result does not rule these out, because they change \emph{which expert gets which token} (or which token lands on which rank), not just which rank holds an expert, and so bypass the EP-invariance constraint. What is ruled out is the weaker default that the EP knob on its own flattens the load. Recent corrective proposals~\cite{han2026llep,laermoe2026,moetuner} report imbalance at a single EP and do not test whether their corrective target is itself EP-sensitive; our flat-across-EP measurement supplies that missing input.

\section{Q2: Mock Data Overestimates Imbalance at Every Batch Size}
\label{sec:mock}

\subsection{The 5$\times$6 Matrix}

\begin{figure}[t]
\centerline{\includegraphics[width=\columnwidth]{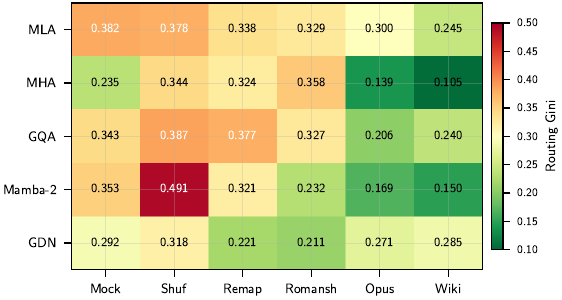}}
\caption{Routing Gini across five architectures and six data conditions, EP=16, GBS=16. Each cell shows the per-rank Gini value (lower is more balanced); colors run from green (balanced) to red (concentrated); bold = best Gini per architecture. Rows show how each architecture responds to the data gradient (left-to-right: synthetic to real-English); columns show how the five architectures route the \emph{same} input differently.}
\label{fig:heatmap}
\end{figure}

The full picture is in Fig.~\ref{fig:heatmap}. Every architecture routes better under wikitext than under mock tokens, with the mock-to-wikitext Gini ratio ranging from $1.02\times$ (GDN, barely moves) to $2.35\times$ (Mamba-2). For the more conservative mock-to-unseen-English (Opus) comparison, the range is $1.08\times$ to $2.09\times$. The conservative end of these ranges should not be over-interpreted: GDN's Gini barely moves because its routing is concentrated to begin with, not because mock and real agree: under both conditions GDN sits at Gini $\sim 0.29$.

\textbf{Operational reading.} A Gini of $0.235$ on MHA-mock means the max-loaded rank receives roughly twice the mean per-rank token count; the same model on wikitext gives Gini $0.105$, which corresponds to a max-loaded rank within about $30\%$ of the mean. A scheduler tuned to the mock distribution provisions tail bandwidth for the former and over-provisions by close to a factor of two against the latter. The same logic applies to expert capacity factors and to topology dimensioning.

\subsection{The Shuffled and Remapped Controls}

The improvement under real text is not because real text gives the router a less-noisy token frequency distribution. The shuffled condition retains the wikitext frequency distribution exactly and breaks only sequential structure, and routes as badly as or worse than mock for every architecture. Mamba-2 is the extreme case: shuffled Gini is $0.491$, $39\%$ worse than its own mock baseline. The SSM amplifies whatever sequential statistics are present, including incoherent ones.

Sequential structure alone is also not enough. The remapped condition keeps wikitext's sequence structure intact but breaks the token-to-embedding map, and routing degrades to mock-tier for every architecture where mock and real differ: MLA's remapped Gini ($0.338$) and GQA's ($0.377$) sit within $0.05$ of their own mock baselines, and MHA's remapped Gini ($0.324$) is in fact $0.09$ \emph{higher} than its mock Gini (GDN's mock and real already agree). The router uses the \emph{correct} learned embedding lookup, not arbitrary sequential context; without it, having real syntax does not help. Bershatsky and Oseledets~\cite{bershatsky2025spatial} previously showed on a single Switch-style MoE that routing uses both semantic and positional information; our shuffled/remapped factorial across five architectures isolates these two axes and adds the embedding-lookup contribution as a third independent factor. This matches OpenMoE's observation that MoE routing is largely context-independent and driven by token identity~\cite{xue2024openmoe}: the remapped condition destroys exactly that token-to-embedding signal, and routing degrades to mock-tier.

The implication for benchmark design is direct. Replacing mock tokens with a token-frequency-matched but sequentially-shuffled real-text corpus (a step a benchmark designer might consider as a ``cheaper real-data'' option) does not approximate real routing behavior. The router needs the sequence and the embeddings together.

\subsection{Routing Imbalance Is Batch-Size-Invariant}
\label{sec:gbs}

\begin{figure}[t]
\centerline{\includegraphics[width=\columnwidth]{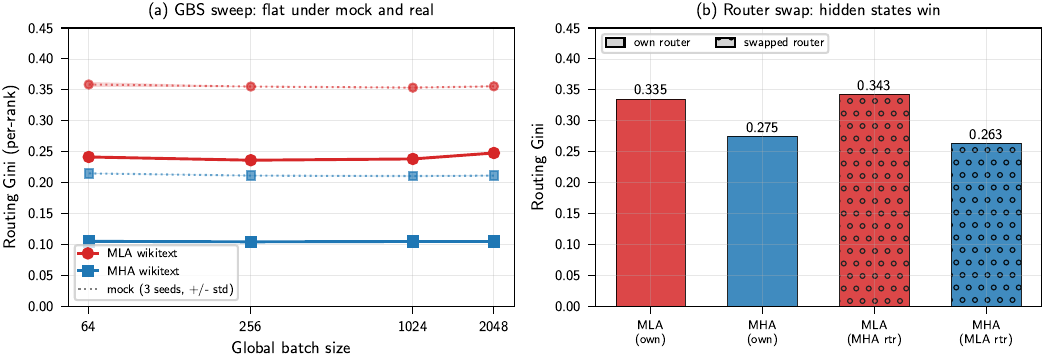}}
\caption{MLA vs MHA, the architecture-controlled pair (only the attention mechanism differs). (a) Per-rank routing Gini vs global batch size under mock (dotted; mean of three seeds, shaded $\pm$1 std) and wikitext-103 (solid). Both flat: the mock-to-real overestimate is a constant factor ($1.4\times$ MLA, $2.0\times$ MHA), not a trend. (b) Router swap: the resulting Gini follows the model body, not the router ($0.335 \to 0.343$ on MLA hidden states, $0.275 \to 0.263$ on MHA).}
\label{fig:mla_mha}
\end{figure}

A second question is whether the mock-vs-real gap grows with batch size. It does not. Fig.~\ref{fig:mla_mha}(a) sweeps GBS from $64$ to $2048$ for the MLA/MHA controlled pair, with the mock condition re-measured over three seeds: per-rank Gini varies by less than $1\%$ for MLA and about $3\%$ for MHA under mock, and by $\leq 3\%$ under wikitext, across the full $32\times$ sweep. The seed-to-seed spread at a fixed batch size (std $\leq 0.004$) is of the same order as the total variation across the sweep. Mock data misleads through its level, not its slope: it holds imbalance at a saturated value that overstates the real-text Gini by a constant factor at every batch size ($1.4\times$ for MLA, $2.0\times$ for MHA here; the per-condition gaps of Fig.~\ref{fig:heatmap} at the probe batch).

The flat result is easy to get wrong with a cheaper measurement protocol. Reading a per-expert Gini at a \emph{single} logged iteration of the same sweep produces an apparent mock-only upward trend of $+17\%$ (MLA) to $+40\%$ (MHA); across three seeds and ten logged iterations per run, however, that statistic has a within-condition standard deviation of $0.05$--$0.06$, larger than the entire apparent slope, and its multi-sample mean is flat ($-0.8\%$ for MLA, $-3.4\%$ for MHA). We spell this out because the single-sample failure mode is exactly the kind of benchmark-calibration error this paper is about: a routing statistic read from one iteration of one run can fabricate a scaling trend that hundreds of iterations refute.

The implication is concrete. A practitioner who concludes from a mock-data profile that ``larger batches make imbalance worse'' has read noise, not workload: routing imbalance at GBS=$2048$ is statistically indistinguishable from GBS=$64$ under both mock and real text, for both 64-expert and 128-expert models. Capacity planning can treat per-rank routing skew as batch-size-invariant and correct for mock data with a single per-model level factor, calibrated under the same released-checkpoint measurement protocol; the factor is a property of the frozen router rather than of the model family, and does not transfer to continued-training regimes (see the adaptation item in \S\ref{sec:discussion}).

The $47\times$ mock-to-real Dirichlet-$\alpha$ gap we reported on Nemotron in \S\ref{sec:intro} is the same effect read in the more uniformity-sensitive metric: mock $\alpha=0.16$ vs unseen-real $\alpha=7.5$. The two distributions sit on opposite sides of ``mostly uniform,'' which is why the ratio is so large in $\alpha$ but more moderate in Gini.

\subsection{Skewing the Synthetic Distribution Makes It Worse}
\label{sec:zipf}

\begin{figure}[t]
\centerline{\includegraphics[width=\columnwidth]{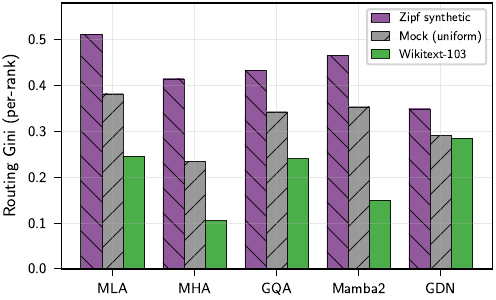}}
\caption{Per-rank routing Gini under skewed-Zipf synthetic tokens, uniform-random mock, and wikitext-103, at the factorial point (EP=16, GBS=16). Zipf token IDs increase imbalance beyond uniform mock for every architecture ($1.2\times$--$1.8\times$ mock, $1.2\times$--$3.9\times$ wikitext): a synthetic generator with more realistic \emph{token frequencies} is a \emph{less} realistic routing proxy.}
\label{fig:zipf}
\end{figure}

If uniform-random tokens overestimate imbalance, a natural repair is to make the synthetic distribution more text-like by skewing it: Tutel evaluates on Zipf-distributed dispatch~\cite{hwang2023tutel}, and real token unigrams are famously Zipfian. We test this directly with a seventh condition: token IDs drawn from a truncated Zipf ($s{=}1.0$) distribution over each model's own vocabulary (top $1\%$ of vocabulary carrying ${\sim}62\%$ of the token mass, versus ${\sim}1\%$ under uniform mock), at the same EP=16, GBS=16 factorial point.

The repair fails in the informative direction (Fig.~\ref{fig:zipf}). Zipf tokens route \emph{worse} than uniform mock for every architecture: per-rank Gini $0.51$ vs $0.38$ mock for MLA, $0.41$ vs $0.24$ for MHA, $0.43$ vs $0.34$ for GQA, $0.47$ vs $0.35$ for Mamba-2, and $0.35$ vs $0.29$ for GDN. Against wikitext the overestimate reaches $3.9\times$ (MHA) and $3.1\times$ (Mamba-2). The mechanism is the one \S\ref{sec:mock} isolates: routing is token-identity-driven, so repeating a small set of token IDs concentrates load on the experts those IDs map to, and skewing the ID distribution amplifies exactly this concentration; real text avoids it not by having balanced unigrams but by pairing its skewed unigrams with learned embeddings and sequence structure the router was trained on. Matching a single marginal statistic of real data (the unigram frequency) while breaking the rest moves the benchmark \emph{away} from real routing behavior, the same lesson as the shuffled and remapped controls. For benchmark design the ranking is unambiguous at this operating point: wikitext $<$ uniform mock $<$ Zipf in routing imbalance, so of the two synthetic defaults the field actually uses, uniform-random is the less wrong one, and neither substitutes for real text.

\section{Q3: Two Emergent Architecture Classes}
\label{sec:tax}

\subsection{The Class Cut}

Reading the columns of Fig.~\ref{fig:heatmap} for the real-data conditions (Romansh, Opus, Wikitext) reveals two extremes and one intermediate case. MHA and Mamba-2 reach Gini $0.105$ and $0.150$ under wikitext (improvement from mock: $2.24\times$ and $2.35\times$ respectively), the lowest absolute imbalance in our matrix and the largest gain from real data. MLA and GDN stay above Gini $0.24$ under English real text, and above $0.21$ everywhere (improvement: $1.56\times$ and $1.02\times$), the highest absolute imbalance and the smallest gain. GQA is the intermediate case: its wikitext Gini ($0.240$) is in the same band as MLA's ($0.245$), but its mock-to-wikitext improvement ratio ($1.43\times$) is closer to MLA's than to MHA's, and its mock Gini ($0.343$) is lower than MLA's mock ($0.382$). It does not match either extreme on both axes simultaneously: MLA and GDN improve less and stay high; MHA and Mamba-2 improve more and drop low; GQA improves moderately and stays moderately concentrated.

We call the two extremes \emph{data-resilient} (MHA, Mamba-2) and \emph{persistently concentrated} (MLA, GDN), and treat GQA as a mixed case. The taxonomy is descriptive, not architectural: we are not claiming the attention mechanism mechanically determines class membership, only that the five architectures separate this way along the routing-balance dimension under real data.

\begin{figure*}[t]
\centerline{\includegraphics[width=\textwidth]{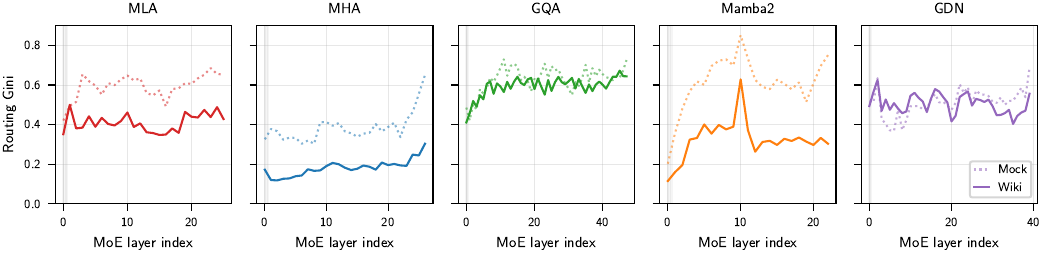}}
\caption{Routing Gini at each MoE layer under wikitext (solid) and mock (dotted) input. Two operational regimes: MHA stays below $0.31$ at every layer and Mamba-2 at or below $0.34$ everywhere except one localized spike ($0.63$ at depth fraction $0.45$, the MoE block preceding an attention layer in the hybrid layout), while MLA, GQA, and GDN sit above $0.35$ at almost every layer. Per-architecture depth profiles and the hash-routing design implication are discussed in \S\ref{sec:tax}.}
\label{fig:perlayer}
\end{figure*}

\subsection{Where the Imbalance Lives in Depth}

The class cut is sharpened, and refined, when read by layer. Fig.~\ref{fig:perlayer} plots Gini per MoE block under wikitext (raw matrices: Appendix~\ref{app:raw}). Two operational regimes are visible: MHA stays below Gini $0.31$ at every layer, Mamba-2 is similarly low for $\sim 90\%$ of its stack with a single sharp spike, and MLA, GQA, and GDN sit above Gini $0.35$ at almost every layer. The depth profile \emph{within} each regime varies more than a single-class story would predict:

\begin{itemize}
\item \textbf{MHA} traces a shallow U-shape: layer-0 Gini $0.17$, dipping to $0.12$ in the early-middle, rising to $0.30$ at the deepest MoE block.
\item \textbf{Mamba-2} starts the lowest of any architecture (layer-0 Gini $0.12$), rises through the early-middle to a single-layer spike of $0.63$ at depth fraction $0.45$ (the MoE block immediately preceding an attention layer in the hybrid stack), then drops back and stabilizes between $0.26$ and $0.34$ for the rest of the stack. The 22 other Mamba-2 MoE blocks are all $\leq 0.40$, so the spike is a localized peak, not a high baseline.
\item \textbf{MLA} is flat-high, fluctuating between $0.35$ and $0.50$.
\item \textbf{GQA} starts at Gini $0.41$, rises sharply to $\sim 0.59$ by depth fraction $0.20$, and stays in the $0.55$--$0.67$ range thereafter.
\item \textbf{GDN} oscillates between $0.40$ and $0.63$ with no clear monotone direction.
\end{itemize}

\emph{Do hybrid models route differently after SSM vs attention?} For both hybrids, the means agree to within noise: Nemotron's SSM-followed MoE blocks sit at wikitext Gini $0.321$ (n=17) vs attention-followed $0.333$ (n=6); Qwen3.5's at $0.505$ (n=30) vs $0.504$ (n=10). The architecture-class signal is a global property of the model, not a function of which mixer precedes a given MoE block; the one nuance is that Mamba-2's depth-0.45 spike lies on an SSM-followed block and drops sharply at the next attention layer, suggesting attention acts as a periodic reset in the SSM's routing-concentration trajectory.

\medskip

This refined picture sits at the intersection of three recent design observations from the model-architecture literature. LASER~\cite{laser} reported, on two MHA-based MoE models (Mixtral and DeepSeek-MoE-16b), a U-shaped gate-score distribution across depth: skewed at the ends, flatter in the middle. Our MHA profile reproduces that U cleanly, but Mamba-2 does not, indicating the U is an MHA-MoE phenomenon rather than a universal one. Ternovtsii and Bilak~\cite{ternovtsii2026geometric} describe a frequency-to-syntax gradient in MoE routing (early layers segregate by word frequency, deeper layers by syntactic category), which implies that early-layer routing is essentially a vocabulary-statistics partition, something a hash function approximates well. DeepSeek-V4~\cite{deepseekv4} acts on that implication by replacing the first few transformer blocks with hash-routed MoE~\cite{roller2021hash}. Our data sharpens the design rule: hash routing is a near-loss-free substitute exactly for the data-resilient class, whose layer-0 Gini under learned routing is already near uniform ($0.17$ for MHA, $0.12$ for Mamba-2). For the persistent class, layer-0 Gini sits at $0.35$--$0.49$: the learned router is doing something stronger than a vocabulary partition at the bottom of the stack, and hash routing would replace that signal with noise without addressing the imbalance source that propagates through the rest of the stack.

The per-layer concentration phenomenon itself is not our contribution; LASER and Geometric Routing each name a piece of it. The architecture-class organization is: the same depth question gives qualitatively different answers across MLA, MHA, GQA, Mamba-2, and GDN, and those answers map onto the two operational bands that Fig.~\ref{fig:heatmap} already identified.

\subsection{Toward a Mechanism}

We are conservative about mechanism, because mechanism claims are easy to overstate. Four observations constrain it.

First, the MLA and MHA models share everything except attention, and they sit at opposite ends of the resilient/persistent split (MLA Gini $0.245$, MHA $0.105$ on wikitext). Attention is doing the work, not the rest of the MoE configuration.

Second, the work is done by the hidden-state geometry the attention produces, not by the router weights operating on it: when we swap the two architectures' routers (DSv2-Lite's router into DS-MoE-16B and vice versa), the resulting routing Gini follows the body of the model, not the router (Fig.~\ref{fig:mla_mha}b). The MHA hidden states route at Gini $\sim 0.27$ regardless of which router scores them; the MLA hidden states route at Gini $\sim 0.34$ either way. The router is a lens whose discrimination is bounded above by the diversity of its input. Fan et al.~\cite{fan2024moedesign} report a complementary invariance: a frozen, randomly-initialized router performs comparably to a learned one, again locating the routing signal in the model body rather than the router parameters.

\begin{figure}[t]
\centerline{\includegraphics[width=\columnwidth]{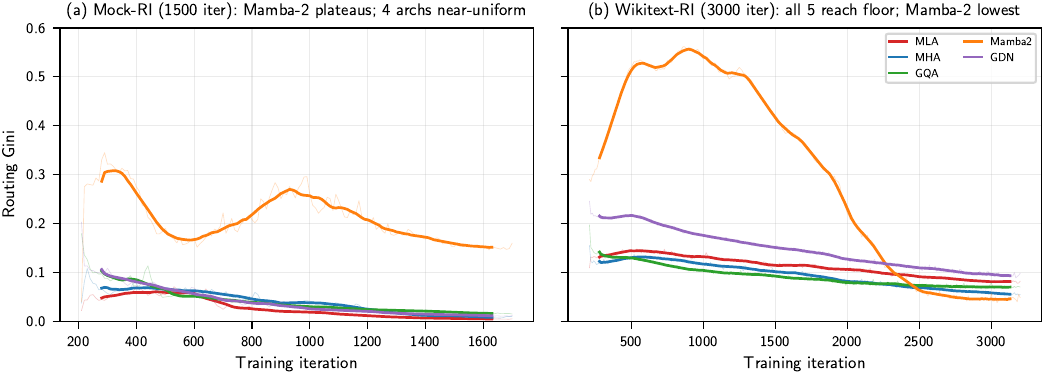}}
\caption{Random-init routing Gini under (a) mock tokens ($200$ warmup + $1500$ measure) and (b) wikitext-103 ($200$ warmup + $3000$ measure, extended to resolve Mamba-2's asymptote). Under mock, four architectures converge to near-uniform while Mamba-2 alone plateaus; under wikitext all five reach stable floors, with Mamba-2 descending from the highest plateau to the lowest asymptote in either panel. Full trajectory details in Appendix~\ref{app:randinit}; both Mamba-2 signatures are consistent with an SSM amplifying the sequential statistics of its input: meaningless mock $\Rightarrow$ noise plateau, coherent text $\Rightarrow$ slow but deep convergence.}
\label{fig:randinit}
\end{figure}

Third, the pretrained two-class taxonomy is not present at random initialization (Fig.~\ref{fig:randinit}). From randomly-initialized weights under \emph{mock} tokens, four of five architectures (MLA, MHA, GQA, GDN) converge to near-uniform routing under standard aux loss (end-of-window Gini $\leq 0.016$); Mamba-2 alone fails to converge, plateauing in the $0.15$--$0.30$ band. Under wikitext, \emph{all five} reach stable floors: the four non-SSM architectures at $0.054$--$0.094$ ($5$--$15\times$ above their mock asymptotes, the structural floor real-text statistics impose on uniformity), and Mamba-2, inverting its mock behavior, descending from the highest plateau ($0.52$--$0.55$, sustained for ${\sim}1000$ iterations) to the \emph{lowest} asymptote of any architecture ($0.047$). The pretrained two-class structure is therefore a property of pretraining, not of the forward pass: at random init under wikitext all five architectures end in a similar low-Gini regime, and the resilient/persistent split is what pretraining-on-real-text produces from that starting point. Appendix~\ref{app:randinit} gives the full trajectories and a measurement-window caution.

Fourth, the Mamba-2 mock-vs-wiki inversion supports a single SSM-amplification mechanism, the same one we posit for its pretrained behavior (largest mock-to-real Gini swing in our 30-cell matrix). The shape difference between the two regimes is itself diagnostic: under mock, Mamba-2 plateaus at $0.15$--$0.30$ with no descent; under wikitext, it plateaus at $0.52$--$0.55$ for $\sim 1000$ iterations before a steep descent to $0.047$. The presence of the descent is the signature of useful sequential statistics that the recurrence can eventually exploit; its absence under mock is the signature of a recurrence amplifying noise. Nemotron's released checkpoint is also the one model whose routing configuration differs along multiple axes (sigmoid scoring and an additive expert-bias term, in addition to the aux loss); we ran a targeted ablation disabling Nemotron's expert-bias term and holding everything else fixed, and the data-sensitivity signature is unchanged ($1.3\%$--$1.4\%$ on Gini, well within noise). This narrows but does not eliminate the routing-config confound. The SSM recurrence remains the leading candidate for what produces both the pretrained data-sensitivity and the random-init mock plateau, and is the only single mechanism that explains both directions of the wikitext-vs-mock asymmetry (worst under mock, best under wikitext) at random init.

The class cut is also robust to dropping Nemotron: in the remaining four architectures (MLA, MHA, GQA, GDN in Table~\ref{tab:models}), all softmax-scored, aux-loss-balanced, no expert-bias term, coefficient $10^{-3}$: MHA still sits at the resilient extreme and MLA, GDN at the persistent extreme, with GQA still the mixed intermediate. The same separation emerges without Nemotron's routing-config differences, so whatever Nemotron is doing differently is not what is producing the cut.

\section{Discussion}
\label{sec:discussion}

\subsection{Implications for the Four System Families}

The findings reweight, but do not refute, each of the four families. \emph{Predictive placement}~\cite{netmoe2025} requires that routing at step $t$ predict routing at step $t{+}1$; our separate temporal-stability analysis (consecutive-iteration Pearson $r$ of expert activation vectors, extending~\cite{yu2025orders} to the SSM and GDN architectures) finds $r > 0.7$ between consecutive steps for MHA, GQA, MLA, and Mamba-2 under wikitext, with GQA reaching $r = 0.93$ and lag-20 still at $r = 0.91$; but GDN collapses to $r \approx 0$ within five steps, leaving history-based placement principled only on very short horizons there. \emph{Adaptive relayout and routing constraints}~\cite{laermoe2026,liu2025c2r} correct the routing distribution itself; the EP-invariance result implies their return-on-cost should be evaluated against an absolute target, with the persistent class as the natural beneficiary. \emph{Hierarchical AlltoAll}~\cite{tran2024ohio} exploits on-node concentration in the send-counts matrix, which the persistent class produces substantially more of than the resilient class, a direct consequence of Gini $0.25$--$0.38$ translating to a max-loaded rank several times the mean. \emph{EP-aware topology}~\cite{wang2024railonly,bernadskiy2025lightmatter} adds bandwidth that helps every architecture uniformly, but the workload model used to dimension it, particularly the assumption that imbalance grows with batch size, needs calibration on real data.

\subsection{For Practitioners and Benchmark Designers}

We have two concrete recommendations.

\emph{For MoE communication benchmarks:} report both mock and a real-data condition. The dual report is enough to expose the single-sample GBS pitfall and the architecture-class differential, and every condition we use is reproducible from public sources with standard tokenization (\S\ref{sec:method}).

\emph{For interconnect and dispatch design:} treat the three regimes separately. For the data-resilient pair (MHA, Mamba-2), real-data routing approaches uniform and the AlltoAll traffic is mostly bandwidth-bound; topology and capacity decisions can use uniform-load models. For the persistent pair (MLA, GDN), expect per-rank Gini above $0.24$ under English real text and $0.29$--$0.38$ under mock, at every batch size and every EP we tested; locality-aware dispatch and topology that takes advantage of concentration on a small subset of inter-node pairs has more to gain here. GQA aligns with the persistent group by absolute Gini ($\sim 0.24$ under wikitext) and should be sized for inter-node concentration on the dispatch side, while remaining the prime candidate for NetMoE-style predictive placement because its routing is the most temporally stable of any architecture we measured. These recommendations are calibrated on our measured range (EP $\leq 32$, up to eight nodes); production-scale validity is discussed below.

\subsection{Validity and Limitations}

\textit{Scale and coverage.} The factorial uses 16 GPUs (4 nodes) and the EP scan reaches 32 GPUs (8 nodes); production MoE training routinely uses 256--4096 GPUs across 32--512 nodes, where AlltoAll spans multiple rail segments and may interact with congestion control we do not exercise. Our EP-invariance claim is supported up to EP=32 across an $\sim 18\times$ NVLink$\to$IB per-GPU bandwidth transition, which we read as evidence that the skew property is bandwidth-regime-independent within the range tested; whether it holds at EP=256 across a multi-rail fabric is open. A larger-scale follow-up (more H100-class or higher-grade nodes to extend EP, and higher per-GPU memory to recover EP=4 for the 128-expert models) would directly broaden the scan. EP=4 is also missing for GQA, Mamba-2, GDN; TP=2 is used for Mamba-2 and GDN at EP=8 (the TP item below isolates the resulting confound to Dirichlet $\alpha$).

\textit{Measurement window and per-iter noise.} Factorial: 10 iterations/cell; EP scan: 200. Per-iter Gini stdev within a fixed (MLA, EP) cell on the EP scan is $0.05$--$0.08$, with cross-EP mean variation of $\leq 5\%$ in max/mean: the cross-EP signal is therefore below the within-EP noise envelope, which strengthens the EP-invariance claim but also bounds the resolution of cross-condition Gini comparisons in the factorial (differences $\lesssim 0.02$ should not be over-read).

\textit{Validation scope.} The Gini--P99 within-architecture correlation ($r \geq 0.76$, \S\ref{sec:ep}) is established on the 30 factorial cells at EP=16 only. We do not extend it across the EP scan because Gini and P99 both co-vary with EP for separate reasons (per-rank-variance vs network-bandwidth), and a cross-EP within-arch correlation would be a confound. Cross-architecture differences in Gini at fixed EP do not translate one-to-one to differences in P99 either, because per-arch baseline latency varies with model size, top-$k$, and message volume (cross-arch $r = 0.69$).

\textit{Tensor parallelism.} Gini- and max/mean-based claims are TP-robust up to TP=2: the scan's own TP=2 points sit within $2.7\%$ and $0.3\%$ of their TP=1 neighbors, and an independent TP=1-vs-2 comparison at EP=16 bounds the difference at $\sim 9\%$ on Gini (Appendix~\ref{app:tp}). TP=4 is different in kind, shifting even the per-expert max/mean by $+18\%$ at fixed EP, so no TP=4 point enters any scan; $\alpha$-based claims are reported with the TP configuration.

\textit{Tokenizer fertility.} On Romansh, tokenizer fertility (tokens per whitespace word) correlates with routing Gini across the five models at $r = 0.94$. On English the five tokenizers have near-identical fertility, so the confound is specific to Romansh-style OOD-language conditions and does not affect the wikitext or Opus results.

\textit{Synthetic-token variants.} Our ``mock'' condition is uniform-random token IDs, the de facto default in communication benchmarks~\cite{pan2025hetermoe}, and \S\ref{sec:zipf} adds the other synthetic default, Zipf-skewed token IDs in the style of Tutel's dispatch patterns~\cite{hwang2023tutel}; both overestimate real-text imbalance, the Zipf variant more. What remains open is whether a generator calibrated on more than the unigram marginal (e.g., matching short $n$-gram statistics under the target model's own embedding table) could reproduce real-text routing cheaply; our shuffled and remapped controls bound how much such a generator must preserve, since frequency alone (shuffled) and sequence structure alone (remapped) each fail.

\textit{Continued-training adaptation.} All conditions profile released checkpoints over short matched windows (5 warmup + 10 measurement iterations in the factorial), so the mock-to-real gap we report is a property of the frozen router. Under continued training the two arms diverge: the auxiliary loss re-balances routing on synthetic tokens, while real text can sustain --- and for some architectures deepen --- expert specialization, so the gap's magnitude, and in at least one case its sign, changes as the router adapts. The per-model level factor of \S\ref{sec:gbs} is therefore a released-checkpoint calibration, not a training-time one; characterizing routing imbalance \emph{during} training is deferred to the extended study noted below.

\textit{Routing mechanism.} All five models use auxiliary-loss-based balancing. Aux-loss-free routing (DeepSeek-V3 bias-shift, Ling-style sigmoid routing~\cite{ling2025}) is the natural follow-up; we do not claim our results extrapolate.

\textit{What we did not measure.} End-to-end training throughput under any system-side intervention; AlltoAll latency vs an idealized interconnect; the cost of a chosen dispatch strategy in production. The natural follow-up is an A/B on locality-aware dispatch for the persistent class, taking the architecture taxonomy as input. An extended study covering production-scale expert parallelism (frontier MoE deployments serve at EP=64--320), routing dynamics under continued training, and trace-driven network simulation --- replaying our measured send-counts matrices through packet- and flit-level simulators across candidate fabric topologies --- is in preparation.

\section{Related Work}
\label{sec:related}

\textit{MoE-EP communication optimization.} FasterMoE~\cite{he2022fastermoe} and Tutel~\cite{hwang2023tutel} introduced dynamic shadowing and adaptive parallelism for MoE training; both are evaluated with synthetic dispatch patterns. Lina~\cite{li2023lina} prioritizes AlltoAll over allreduce; Janus~\cite{liu2023janus} moves experts instead of tokens; SmartMoE~\cite{zhai2023smartmoe} co-locates hot and cold experts; NetMoE~\cite{netmoe2025} exploits routing locality via integer-programmed sample placement; LAER-MoE~\cite{laermoe2026} relays out experts adaptively at training time; C2R~\cite{liu2025c2r} adds routing-side constraints to cut AlltoAll volume.

\textit{Interconnect and topology.} Rail-only~\cite{wang2024railonly} eliminates the spine layer of fat-tree topologies for LLM training, citing LLM training's sparse-pattern communication. Lightmatter's 3D-CPO~\cite{bernadskiy2025lightmatter} models a $2.7\times$ MoE training speedup from scaled-up photonic interconnect, motivated explicitly by AlltoAll dominating forward-pass latency. UCC~\cite{aderholdt2024ucc} provides a unified collective-library substrate for CPU/GPU/DPU.

\textit{Collective optimization.} OHIO~\cite{tran2024ohio} optimizes RDMA MPI\_Alltoall scalability through hierarchical aggregation and intra/inter-node overlap. The Chakra representation~\cite{yoo2024chakra} standardizes deep-learning collective traces and is part of the ASTRA-sim simulator ecosystem; our per-layer, per-step \texttt{input\_splits} logs across the $5{\times}6$ matrix convert directly to this format, and we release them so that simulation studies can replay measured, architecture-specific routing in place of synthetic dispatch.

\textit{Communication characterization for distributed transformers.} Anthony et al.~\cite{anthony2024demystifying} characterized DP, TP, and PP communication for dense GPT-NeoX models up to 13B parameters on Frontier, explicitly leaving expert parallelism for future work. Alnaasan et al.~\cite{alnaasan2024peft} did the analogous characterization for parameter-efficient fine-tuning. Our work is the MoE-EP chapter of the same characterization series.

\textit{MoE routing characterization across architectures.} LASER~\cite{laser} characterized per-layer gate-score distributions on two MHA-based models and observed the U-shape our resilient-class measurement is consistent with; LPR~\cite{latentproto2025} reports baseline routing Gini for three models under FineWeb/C4. Both characterize a subset of attention families and a single data condition; our matrix extends both axes. FlexMoE~\cite{nie2023flexmoe} characterizes imbalance across BERT, GPT, and Swin backbones, but at a fixed parallelism rather than across an EP scan. Two model-side studies bear directly on our mock-vs-real result: OpenMoE~\cite{xue2024openmoe} finds routing largely context-independent and token-ID-driven, and OLMoE~\cite{muennighoff2024olmoe} analyzes vocabulary-specialized routing; both predict that removing token-identity information (our remapped and mock conditions) degrades routing, as we observe. Multilingual routing work~\cite{multilingmoe2025} is complementary to our Romansh and tokenizer-fertility observations. MoE-Inference-Bench~\cite{moeinferencebench2025} flags synthetic-to-real transferability as open, the question \S\ref{sec:mock} answers for routing imbalance.

\textit{MoE load imbalance from the model side.} Switch Transformers~\cite{fedus2022switch} and ST-MoE~\cite{zoph2022stmoe} introduced the auxiliary-loss family that our five models all use (with the Nemotron variant adding an expert-bias term). Expert Choice routing~\cite{zhou2022expertchoice} is an alternative balancing mechanism that we did not test. DeepSeek's expert parallelism load balancer~\cite{deepseek2025eplb} (EPLB) periodically remaps experts based on observed workload in production.

\section{Conclusion}
\label{sec:conclusion}

Routing imbalance in MoE expert parallelism is not a scaling problem; it is a model-and-data property the system layer can localize, predict, and partially exploit, but cannot make uniform by EP scaling alone. The synthetic-token benchmarks used to design and evaluate MoE-aware interconnect proposals miss this in two ways: they overestimate imbalance by a constant factor that persists across batch sizes, and skewing the synthetic token distribution toward realism (Zipf) widens rather than closes the gap. The five architectures we measured separate into a data-resilient pair (MHA, Mamba-2) where real data drives routing near-uniform, a persistently concentrated pair (MLA, GDN) where Gini stays high regardless of input or EP, and an intermediate case (GQA); the three regimes have qualitatively different load profiles for interconnect and dispatch design. We offer the architecture taxonomy and the $5{\times}6$ characterization as a calibration target for the interconnect community.

\clearpage

\appendices

\section{Load-Balancing Notation in Table~\ref{tab:models}}
\label{app:notation}

``aux'' refers to the standard auxiliary load-balancing loss~\cite{fedus2022switch}, a small penalty added to the training objective that nudges the router toward balanced expert usage across each minibatch; it is the workhorse mechanism MoE training uses to keep no single expert idle or saturated. ``seq-aux'' is DeepSeek-V2's per-sequence variant: the same penalty applied to each input sequence individually rather than the whole minibatch, encouraging balance within each input as well as across them. ``+bias'' on Nemotron is an additive scalar attached to each expert's router score, learned to drift in response to historical load and complementing the aux-loss penalty. The aux-loss coefficient sets the strength of the penalty relative to the main language-modeling loss.

\section{Topology Footprint of the EP Scan}
\label{app:topo}

The four EP points span the cluster's full bandwidth hierarchy. EP=4 fits within a single node, so every AlltoAll edge is NVLink (no IB hop). EP=8 uses two nodes, so roughly half the partner ranks of each GPU are intra-node (NVLink) and half are inter-node (IB). EP=16 uses four nodes, so 3 of every rank's 15 partners are intra-node and 12 cross IB. EP=32 uses eight nodes, so only 3 of 31 partners are intra-node; the AlltoAll is IB-dominated. Per-GPU bandwidth differs by roughly $18\times$ between intra-node NVLink ($\sim 450$\,GB/s aggregate per GPU, H100 NVLink4 unidirectional) and NDR200 IB ($\sim 25$\,GB/s per GPU, one HCA per GPU at 200\,Gb/s) on Helma, so the EP scan is not just a software knob but a sweep across two qualitatively different bandwidth regimes.

\section{TP Sensitivity and the EP=4 Point}
\label{app:tp}

For GQA, Mamba-2, and GDN, EP=4 would require 32 experts per GPU at $d{=}2048$--$2688$, which does not fit in the H100's 94\,GB below TP=4. We did run the missing EP=4 point at TP=4 and found it cannot be mixed into the scan: a control at fixed EP=8 shows the TP=4 layout itself shifts GQA's per-expert max/mean by $+18\%$ over TP=1 (28.0 vs 23.8; two seeds at EP=4 agree within $\pm 3\%$). The TP=2 points the scan does use are validated by the scan itself: Mamba-2's EP=8 TP=2 point sits within $2.7\%$ of its TP=1 neighbors and GDN's within $0.3\%$, on lines whose total variation defines the flatness claim. Closing the EP=4 gap for the 128-expert models therefore requires larger-memory GPUs at TP$\leq$2, not more tensor parallelism.

A separate validity check compares independent TP=1 and TP=2 runs at EP=16: per-rank Gini differs by up to $\sim 9\%$ (a bound that includes run-to-run variation) and Dirichlet $\alpha$ by $\sim 20\%$, both small against the $1.4\times$--$2.35\times$ cross-condition effects this paper reports.

\section{Random-Init Trajectory Details}
\label{app:randinit}

Under mock tokens (Fig.~\ref{fig:randinit}a, $200$ warmup + $1500$ measure), MLA, MHA, GQA, and GDN collapse together toward near-uniform under standard aux loss (end-of-window Gini $\leq 0.016$), the convergence story the textbook expectation predicts. Mamba-2 alone fails to converge, plateauing in the $0.15$--$0.30$ band for the full window (peak Gini $0.345$ at iter $290$, end-of-window $0.151$).

On wikitext-103 (Fig.~\ref{fig:randinit}b, $200$ warmup + $3000$ measure, a window doubled because Mamba-2's wikitext trajectory does not finish converging in a mock-matched $1500$-step budget), all five architectures reach a stable floor by iteration $\sim 2500$. The four non-SSM architectures asymptote at $0.054$ (MHA), $0.070$ (GQA), $0.081$ (MLA), and $0.094$ (GDN). Mamba-2's trajectory rises from $0.32$ to a sustained $\sim 1000$-iteration plateau at $0.52$--$0.55$ (peak $0.547$ at iter $910$; the only trajectory whose Gini rises substantially before falling), descends steeply between iterations $1500$ and $2500$, and settles at $0.047$.

Window length matters for this measurement: a short window (e.g., sampling iterations $21$--$69$) catches only the rising edge of Mamba-2's plateau and would suggest a class distinction at random init that the resolved asymptotes do not support.

\section{The Raw Send-Counts Matrices}
\label{app:raw}

\begin{figure*}[!t]
\centerline{\includegraphics[width=0.86\textwidth]{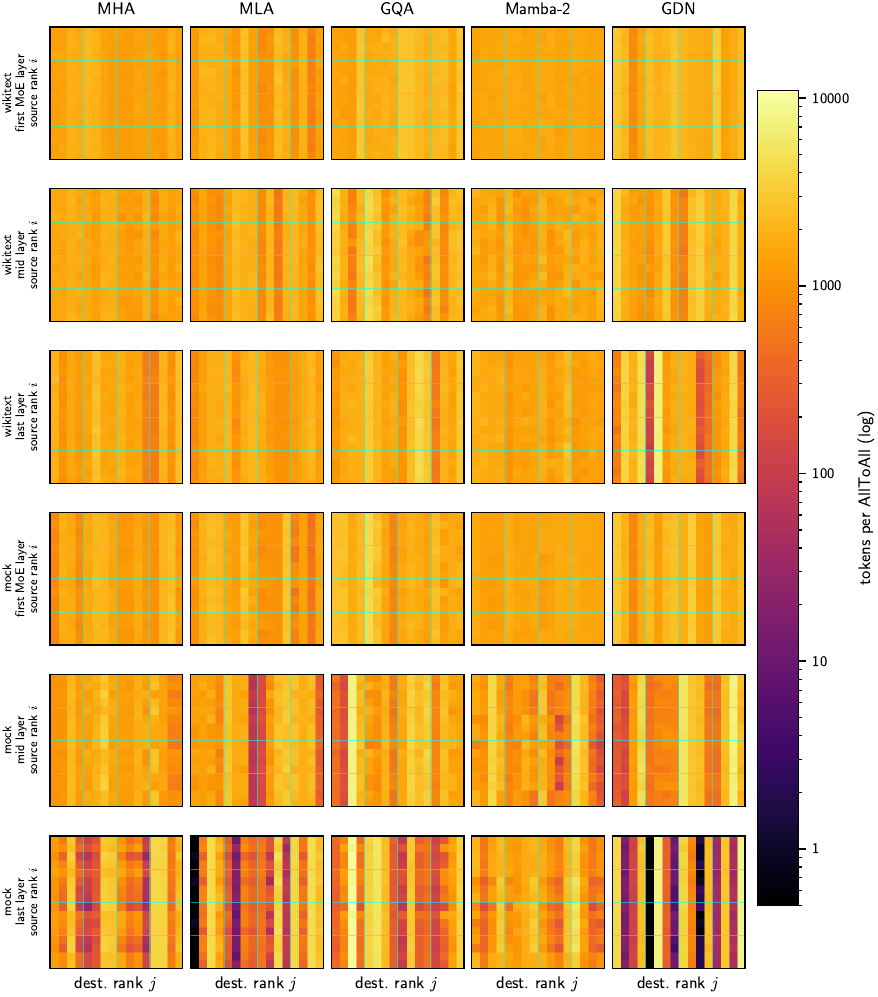}}
\caption{Single forward-dispatch AllToAll send-counts matrices $S$ at the paper's factorial operating point (EP=16, GBS=16, released checkpoints; row $i$ = tokens rank $i$ sends, column $j$ = tokens rank $j$ receives; cyan grid = 4-GPU node boundaries). Rows 1--3: first, middle, and last MoE layer under wikitext-103; rows 4--6: the same layers under mock. The structure everywhere is \emph{vertical stripes} --- every source rank converges on the same hot destination ranks, not the block-diagonal, node-local pattern an idealized placement-friendly workload would show. A hot \emph{column} is why AlltoAll completion tracks $\max_j \sum_i S_{ij}$ (\S\ref{sec:bg:moe-ep}) and why re-assigning experts to ranks cannot flatten it (\S\ref{sec:ep}). Each layer has its own hot-column set, and the real-text depth structure follows Fig.~\ref{fig:perlayer}. Mock nearly agrees with real text at the first MoE layer, consistent with early-layer routing approximating a vocabulary partition (\S\ref{sec:tax}), but increasingly overstates concentration with depth (at the last layer up to $4.7\times$ vs $2.1\times$ max/mean column, GQA), including near-starved cold columns that real text never produces.}
\label{fig:ep16stripes}
\end{figure*}

\begin{figure*}[!t]
\centerline{\includegraphics[width=\textwidth]{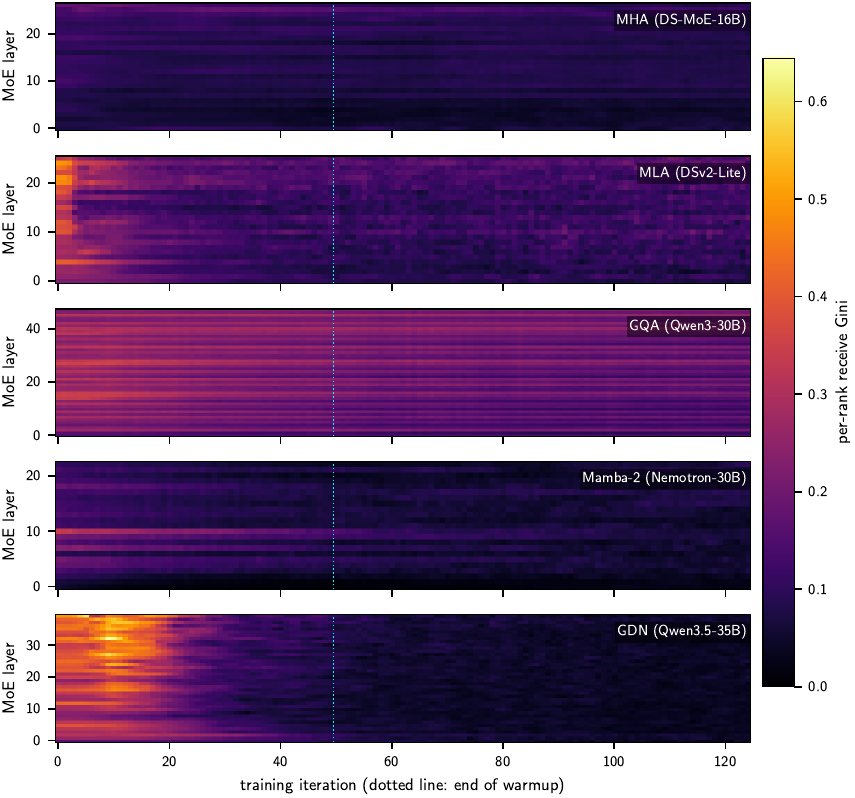}}
\caption{Per-layer, per-iteration routing imbalance under wikitext-103 at the factorial EP=16 operating point, measured with the EP scan's own protocol (GBS=32; per-rank receive Gini of each forward dispatch AllToAll; dotted line marks the end of the 50-iteration warmup; x-axis cropped to the span common to all five cells). The depth profiles of Fig.~\ref{fig:perlayer} appear as horizontal stripes that persist across the window --- most prominently GQA's stable high-Gini bands --- while the auxiliary loss re-balances the concentration the released checkpoints start with, at architecture-specific rates: fastest for GDN, slower for MLA, with MHA and Mamba-2 low throughout. The within-window drift is why the EP scan of \S\ref{sec:method:scans} compares identical iteration windows at every EP point (the mock EP=32 scan windows drift the same way), and the continued-training re-balancing is why the mock-to-real level factor of \S\ref{sec:gbs} is a released-checkpoint calibration that does not transfer to continued-training regimes (adaptation item, \S\ref{sec:discussion}).}
\label{fig:wikiheat}
\end{figure*}

Fig.~\ref{fig:ep16stripes} shows the raw object every claim in this paper summarizes: the per-AllToAll send-counts matrix, under the production-relevant real-text condition first and the benchmark-default mock for contrast. Its structure is column-striped incast, not block-diagonal locality, under both conditions; its intensity is condition-dependent. Fig.~\ref{fig:wikiheat} extends the same real-text condition across training time: the per-layer routing Gini of every forward dispatch AllToAll over a scan-protocol window, showing the within-window drift that motivates the matched-window design and the continued-training adaptation that scopes the paper's level-factor calibration to released checkpoints.

\newpage

\section*{Acknowledgment}
This work has been funded by the Free State of Bavaria in the DSgenAI project (Grant Nr.: RMF-SG20-3410-2-18-4). The authors gratefully acknowledge the scientific support and HPC resources provided by the Erlangen National High Performance Computing Center (NHR@FAU) of the Friedrich-Alexander-Universität Erlangen-Nürnberg (FAU). The hardware is funded by the German Research Foundation (DFG).


\bibliographystyle{IEEEtran}
\bibliography{references}

\end{document}